\documentclass[letterpaper, 10 pt, conference]{ieeeconf}  

\IEEEoverridecommandlockouts                              

\usepackage{graphicx} 
\usepackage{amsmath} 
\usepackage{amssymb}  
\usepackage{algpseudocode}
\usepackage{algorithm}

\newcommand{\trn}{^T}
\newcommand{\Q}{\mathcal Q}
\newcommand{\R}{\mathbb R}
\newcommand{\Sw}{\mathcal S}
\newcommand{\Hm}{\mathcal H}
\DeclareMathOperator{\rank}{rank}
\usepackage{tikz}
\usepackage{tikz-qtree}
\usetikzlibrary{trees}

\title{\LARGE \bf
Real-Time Planning with Primitives for\\ Dynamic Walking over Uneven Terrain
}

\author{Ian R. Manchester and Jack Umenberger
\thanks{This work was supported by the Australian Research Council.}
\thanks{The authors are with the Australian Centre for Field Robotics (ACFR), Department of Aerospace, Mechanical and Mechatronic Engineering, University of Sydney, NSW, 2006, Australia.
        {\tt\small ian.manchester@sydney.edu.au}}%
}

\begin{document}

\maketitle
\thispagestyle{empty}
\pagestyle{empty}

\begin{abstract}
We present an algorithm for receding-horizon motion planning using a finite family of motion primitives for underactuated dynamic walking over uneven terrain. The motion primitives are defined as virtual holonomic constraints, and the special structure of underactuated mechanical systems operating subject to virtual constraints is used to construct closed-form solutions and a special binary search tree that dramatically speed up motion planning.  We propose a greedy depth-first search and discuss improvement using energy-based heuristics. The resulting algorithm can plan several footsteps ahead in a fraction of a second for both the compass-gait walker and a planar 7-Degree-of-freedom/five-link walker.
\end{abstract}

\section{INTRODUCTION}

Passive dynamic walkers are an inspiring topic of research because of the way in which a physical mechanism -- an arrangement of masses, joints, springs, etc -- can create remarkably life-like walking motions without actuation or deliberate motion planning \cite{mcgeer1990passive}. Underactuated dynamic walkers -- a.k.a. limit cycle walkers --  add minimal actuation but retain many of the same properties \cite{collins2005efficient}.

Despite their aesthetic appeal, the practical utility of dynamic walkers is severely limited by two facts: firstly, the motions are usually limited to periodic walking on flat ground or down a shallow slope, and secondly, these motions typically have extremely small regions of attraction. 

A number active control techniques have been developed to improve the stability of periodic walking motions,  e.g., virtual constraints \cite{grizzle2001asymptotically}, \cite{westervelt2003hybrid}, \cite{westervelt2007feedback}, controlled Lagrangians \cite{spong2005controlled}, and transverse linearization \cite{freidovich2008stability}. Feedback control with a periodic target gait can also achieve impressive results on uneven terrain, as evidenced by the famous BigDog robot and its siblings \cite{raibert2008bigdog}. In \cite{park2013finite} the virtual constraints methodology was expanded to include ``reflex'' motions when unexpected terrain variations are detected, and \cite{byl2009metastable} analysed  stability of walking on stochastically generated terrain. Nevertheless, it is clear that for large terrain variations, some kind of motion planning based on the terrain ahead of the robot will be beneficial. Indeed, numerical results on the compass gait walker in \cite{byl2008approximate} suggest that even being able to plan a few steps ahead can help significantly.

\begin{figure}
\begin{center}
\includegraphics[width=0.8\columnwidth]{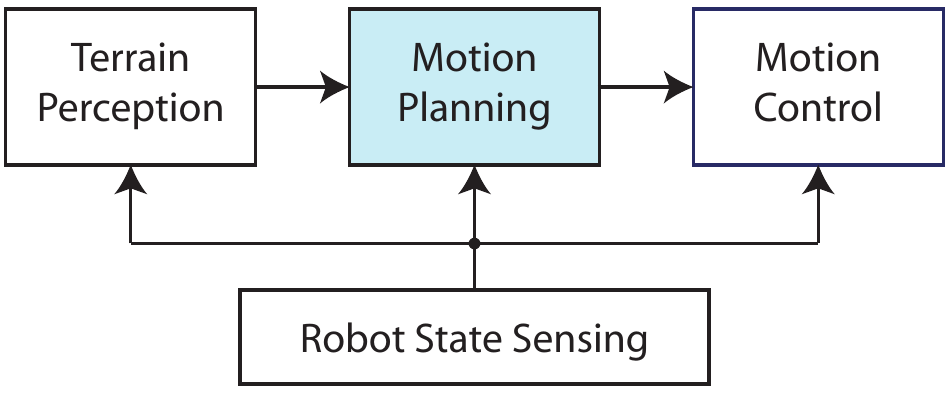}
\caption{Possible organization of perception and control of a walking robot. In this paper we consider the problem of motion planning.}
\label{fig:blockdiag}
\end{center}
\end{figure}

In this paper, we assume the robot control system has a task breakdown of the form in Figure \ref{fig:blockdiag}. In real time, the robot must sense the terrain ahead, continuously re-plan a feasible motion over several footsteps, perhaps a few times per second, and use active control to stabilize this motion with a sampling frequency between 100Hz and 1kHz.  In \cite{manchester2011stable} it was shown that the transverse linearization can be used for active control of such motions using a receding horizon (model predictive control) approach. The purpose of the present paper is to address the motion planning problem: that is, given knowledge of the terrain ahead and the dynamic state of the robot, rapidly produce a feasible walking motion.

Many of the classic algorithms for robot motion planning focus on finding collision-free paths in configuration space \cite{spong2008robot}, \cite{lavalle2006planning}, since a fully actuated industrial robot can track any continuous configuration space path, and the dynamics only affect the timing with which it is tracked. Planning for Asimo-like walking robots usually supplements these algorithms with an extra constraint to keep the centre of pressure -- the ``zero moment point'' -- inside the convex hull of foot contact points, so the robot remains fully actuated (see, e.g., \cite{chestnutt2005footstep}).

Underactuated robots have differential constraints that typically render most configuration-space paths infeasible. The most direct approach to motion planning is to pose it as an optimization problem over state and control trajectories, and apply the methods of nonlinear programming. Unfortunately, walking robot motion planning problems are typically high-dimensional, nonsmooth, and nonconvex. Significant progress has been made recently on this front, see, e.g., \cite{erez2012trajectory}, \cite{posa2013direct}, however with current algorithms it can still take several minutes to compute a motion on a multi-core desktop computer.

The use of primitives converts an infinite-dimensional optimization over states and inputs to the combinatorial problem of selecting a finite sequence from a fixed library of primitives. Complete search over the tree of possible sequences obviously has computational complexity that grows exponentially in the sequence length, so the principle aim in algorithm development is to reduce the search via heuristics. A breakthrough technique was the RRT, which aims to expand the tree in the direction of randomly sampled states \cite{lavalle2001randomized}, \cite{lavalle2006planning}, though for highly dynamic systems this can be difficult. In \cite{frazzoli2002real} motions of a helicopter were planned by augmenting the RRT special primitives and a lower bound on the cost-to-go function. In \cite{shkolnik2011bounding}, bounding motions over rough terrain were planned for the LittleDog robot using an RRT with reachability-guided sampling of action primitives. Again, the planning times on a multi-core desktop were several minutes, because evaluation of each primitive required numerically solving a high-dimensional nonlinear differential equation.


In this paper, we suggest that using virtual constraints as primitives offers three principle advantages for motion planning: (i) they correspond to configuration paths, and can therefore inherit much of the research on planning collision free paths for fully actuated robots; (ii) a partial closed-form solution of the hybrid zero dynamics greatly reduces the computational effort in checking dynamic feasibility of paths and computing total mechanical energy; (iii) the affine structure of the solution allows intelligent ordering of primitives greatly reducing the {\em number} of primitives that must be checked for dynamic feasibility.

We propose a ``greedy'' best-first-search algorithm and discuss improvements based on an energy heuristic. These algorithms are evaluated via simulation on the compass-gait biped and a seven degree-of-freedom/four actuator biped 
\cite{plestan2003stable} which is modelled on the Rabbit robot of CNRS \cite{chevallereau2003rabbit}.

\section{Underactuated Walking Robots}

In this paper we focus on saggital-plane models of underactuated walking robots, however the methods we propose could be extended to three-dimensional models using the reduction method in \cite{gregg2012control} and to some other classes of system such as brachiating robots. 

We consider robots modelled with the methods of classical mechanics. The {\em configuration} is a set of $n$ coordinates $q \in \Q$, an $n$-dimensional smooth manifold (possibly with a boundary) that specifies the set of feasible geometric arrangements of the robot. The state of the system is $(q, \dot q) \in T\Q$, where $T\Q$ is the tangent bundle of $\Q$. If a robot has an $n$-dimensional configuration manifold, its state space is $2n$-dimensional.

We will make the common assumption \cite{westervelt2007feedback} that the biped robot has a {\em stance leg} which does not slip during continuous phases and can be considered ``pinned''. Hence we will consider $q$ to be made up of the remaining links, which we assume form an open chain. At impact, the other leg -- the {\em swing leg} -- becomes the new stance leg, and vice-versa.

The continuous dynamics are derived via a Lagrangian of the form $L(q,\dot q) := \frac{1}{2}\dot q\trn M(q)\dot q-V(q)$, where $V(q)$ gives potential energy, and $M(q)$ is a symmetric positive-definite mass/inertia matrix. The Euler-Lagrange equation:
\begin{equation}\label{eq:EulerLagrange}
\frac{d}{dt} \frac{\partial L(q,\dot q)}{\partial \dot q} = \frac{\partial L(q,\dot q)}{\partial q} + B(q)u,
\end{equation}
results in a differential equation of the form \cite{spong2008robot}
\begin{equation}\label{eq:dynamics}
M(q(t))\ddot q(t) + C(q(t), \dot q(t))\dot q(t) + G(q(t)) = B(q)u,
\end{equation}
where $u$ is a vector signal of actuator forces and torques, $C(q, \dot q)$ consists of Coriolis and centrifugal terms, and $G(q)$ is the gradient of the potential energy field. For this paper, we assume that friction and disturbance forces are negligible or can be counteracted by feedback control.

A model of a walking robot typically also employs an {\em impact map}: a discrete mapping that represents sudden changes of velocities that occur during collisions, e.g. when the swing leg touches down to the walking surface and becomes the new stance leg. In this paper, we assume that collisions are purely inelastic and the impact map has the the following form \cite{hurmuzlu1994rigid}:
\begin{align}
q(t^+) &= Rq(t^-),\label{eq:impact1}\\
\dot q(t^+) &= R\Delta(q(t^-))\dot q(t^-),\label{eq:impact2}
\end{align}
which occurs when $q(t^{-})\in \Sw$, a switching surface in configuration space. For the systems we consider the matrix $R$ represents simply a relabelling of coordinates, e.g. redefining the stance leg as the swing leg and vice-versa. Note also that the mapping for $\dot q(t^+)$ is linear with respect to $\dot q(t^-)$. For simplicity, in this paper we restrict consideration to models with a single impact per footstep, but extension to systems with any finite number of impacts per footstep (e.g. due to knee locking) is straightforward.

As mentioned in the introduction, for {\em fully actuated} robots, i.e. those for which $\rank B(q)=n$ for all $q\in\Q$, every configuration path is dynamically feasible, and a common strategy in motion planning for fully actuated robots is to first plan a collision-free path through configuration space, and then plan or control a dynamically feasible time parameterization \cite{lavalle2006planning}.



Unfortunately this is not generally possible for underactuated systems: if $B(q)$ has fewer than $n$ linearly independent entries, then there exists collision-free trajectories $q^\star(\cdot)$ that are infeasible. The special case of {\em underactuation degree one} systems, i.e. those that have $n-1$ independent actuators, presents very useful extra structure and includes several practically important systems, including several common models for walking robots. In particular, under some mild assumptions, a trajectory  $q^\star:[0, S] \rightarrow \mathcal Q$ is {\em quasi-feasible} in the following sense: it is possible to control the system so that for all $t$, $q(t) = q^\star(\tau)$ for some $\tau$, however the {\em dynamics} of $\tau$ cannot be directly controlled, but are fixed by the so-called {\em zero dynamics}. Therefore, when planning a path through configuration space one must also take into account the zero dynamics.

\section{Problem Statement}

The receding-horizon planning-with-primitives problem consists of two parts: (i) construction of a rich library of useful motion primitives, and (ii) an algorithm for choosing a primitive sequence in real time. It is assumed that the algorithm has as inputs the state of the robot $q, \dot q$, as well as a height-map of the terrain ahead.

For the particular problems we consider, the robot is a planar walker and the terrain data is a one-dimensional height-map over a finite interval, i.e. a function $h:[x_1, x_2]\rightarrow \R$. In practice, this may come from fused measurements of a laser scanner or image sensors. We do not address the problem of terrain sensing in this paper, however it is a well-studied problem over the last few decades, see, e.g., \cite{krotkov1994terrain, triebel2006multi, vasudevan2009gaussian}.

Formally, a set of primitives can be modelled as a finite alphabet $\mathcal P$. Then a $k$-step motion-planning algorithm is a function $T\Q\times \Hm \rightarrow \mathcal P^k$, where $\Hm$ is a suitable function space for height maps. In a receding horizon architecture, the first primitive in the sequence is sent to the motion control module to be regulated, and then the process repeats.

\section{Virtual Holonomic Constraints as Motion Primitives}

The use of virtual constraints for periodic dynamic walking was introduced in \cite{grizzle2001asymptotically} and further studied in \cite{westervelt2003hybrid}, \cite{westervelt2007feedback}, \cite{manchester2011stable}, \cite{park2013finite}. The idea is that a single generalized coordinate, denoted $\theta$ is chosen as a ``phase variable'' and all other generalized coordinates are synchronized to functions of $\theta$. This is reminiscent of a 19th-century mechanical horse, in which a single motor drives many joints through clever mechanical linkages to create the illusion of a natural walking motion. The difference is that virtual constraints are enforced by feedback control rather than physical linkages. 

Let us assume that $\theta$ is monotonically increasing over an interval $[\theta_0, \theta_f]$ during the planned trajectory, then one can construct functions $\phi_i(\theta)$ for the generalized coordinates so that the planned motion satisfies:
\begin{equation}\label{eq:vc}
q_i(t) = \phi_i(\theta(t)), \ i = 1, 2, ..., n.
\end{equation}
The above condition is referred to as a {\em virtual constraint}. If the constraints are perfectly regulated, then clearly the following velocity relations also hold:
\begin{equation}\label{eq:vcdot}
\dot q_i(t) = \frac{\partial\phi_i(\theta(t))}{\partial \theta}\dot\theta, \ i = 1, 2, ..., n.
\end{equation}
We define the vector function $\Phi:[\theta_0, \theta_f]\rightarrow \Q$ as $\Phi(\theta) = [\phi_1(\theta), \phi_2(\theta), \hdots, \phi_n(\theta)]\trn$, and use the notation $\Phi'(\theta) = \left[\frac{\partial \phi_1(\theta)}{\partial \theta}, \frac{\partial \phi_2(\theta)}{\partial \theta}, \hdots, \frac{\partial \phi_n(\theta)}{\partial \theta}\right]\trn$ and similarly for second derivatives $\Phi''(\theta)$.

In general, $\theta$ can always be chosen as path length along a trajectory. However, for several common classes of robot it is convenient to take $\theta$ as the unactuated coordinate, e.g. the ankle angle in the compass gait walker. In that case, we can choose a parameterization with $q_i$, $i = 1, ..., n-1$ the directly actuated coordinates and $q_n = \theta$ the unactuated coordinate. In general, on the target trajectory $q, \theta$ provide excessive coordinates for the system, so this representation can always be achieved by local change of coordinates and dropping one of the elements of $q$. 

The $2n-2$ conditions given in \eqref{eq:vc} and \eqref{eq:vcdot} constrain the $2n$-dimensional state space of the system to a two-dimensional ``zero dynamics'' manifold parameterized by $\theta, \dot\theta$. It is straightforward to show that the zero dynamics have the following form during continuous phases \cite{shiriaev2005constructive}
\begin{equation}\label{eq:abg}
\alpha(\theta)\ddot \theta(t) + \beta(\theta)\dot\theta^2+\gamma(\theta) = 0,
\end{equation}
where
\begin{align}
\alpha(\theta) &= B^\perp(\Phi(\theta))M(\Phi(\theta))\Phi'(\theta),\label{eq:abga}\\
\beta(\theta) & = B^\perp(\Phi(\theta))[M(\Phi(\theta))\Phi''(\theta)+C(\Phi(\theta), \Phi'(\theta))\Phi'(\theta)],\notag\\
\gamma(\theta) & = B^\perp(\Phi(\theta))G(\Phi(\theta)),\label{eq:abgg}
\end{align}
where $B^\perp(q)$ is a row vector satisfying $B^\perp(q)B(q)=0$.

In \cite{westervelt2003hybrid} conditions were derived for periodic cycles that ensure that the impact does not shift the system off the virtual constraint manifold. That is, if $q^- = \Phi(\theta^-)$ and $q^- = \Phi'(\theta^-)\dot\theta^-$ then the output of the impact map \eqref{eq:impact1}, \eqref{eq:impact2} satisfies $q^- = \Phi(\theta^-)$ and $q^- = \Phi'(\theta^-)\dot\theta^-$. This is referred to as {\em invariance} of the hybrid zero dynamics. It is straightforward to extend this to the non-periodic case, see \cite{manchester2011stable}.

Assuming the hybrid zero dynamics are invariant, the impact dynamics reduce to a map of the following form for $\theta, \dot\theta$.
\begin{equation}\label{eq:thetaImpact}
\theta^+ = \kappa, \dot\theta^+ = \delta\dot\theta^-, 
\end{equation}
imposed when $\theta = \theta_f$.
Note since we assume impact conditions are defined by configurations alone, $\theta^+$ is a fixed value for each virtual constraint.

\subsection{Partial Closed-Form Solutions for Velocity and Energy}

A useful property of virtual constraints is the fact that a partial closed-form solution of the reduced dynamics \eqref{eq:abg} can be computed {\em off-line}. The solution is {\em partial} in the sense that we do not obtain solutions of $\theta(t), \dot\theta(t)$ as functions of time, but instead we obtain an expression for $\dot\theta$ as a function of $\theta$.

This property has been used before for stabilizing control design \cite{shiriaev2005constructive}, \cite{manchester2011stable}, and searching for periodic cycles for passive walkers \cite{freidovich2009passive}. The special structure we take advantage of is that $\dot\theta$ only enters \eqref{eq:abg} as a squared term. Indeed, the chain rule gives
$
\frac{d}{dt}\dot\theta(t)^2 = 2\dot\theta(t)\ddot\theta(t).
$
Since $\theta$ is monotonic, it can be used as a new independent variable, i.e. the time $t$ can be written as a function of $\theta$, giving 
\[
\frac{d}{d\theta}\dot\theta(t(\theta))^2 = \frac{d}{dt}\dot\theta(t(\theta))^2\frac{dt}{d\theta} = 2\ddot\theta(t(\theta)).
\]
For virtually constrained systems, this can be combined with \eqref{eq:abg} to give
\begin{equation}\label{eq:linsys}
\frac{d}{d\theta}\dot\theta(\theta)^2 =-2\frac{\beta(\theta)}{\alpha(\theta)}\dot\theta(\theta)^2-2\frac{\gamma(\theta)}{\alpha(\theta)}.
\end{equation}
Let us assume for the moment that $\alpha(\theta)\ne 0$ for $\theta\in[\theta_0, \theta_f]$, we will return to this assumption in the next subsection. Since \eqref{eq:linsys} is a linear $\theta$-varying differential equation, it can be  solved over any interval by numerical quadrature, to give an expression of the form:
\begin{equation}\label{eq:GammaPsi}
\dot\theta(\theta)^2 =\Gamma(\theta,\theta_0)\dot\theta_0^2+\Psi(\theta,\theta_0),
\end{equation}
where the scalar functions $\Gamma, \Psi$ can be precomputed for particular intervals $[\theta_0, \theta]$. This implies that on-line computation of $\dot\theta(\theta)^2$ from $\dot\theta(\theta_0)^2$  requires only a single scalar multiplication and a single scalar addition.

Since the impact map for $\dot\theta$ given in \eqref{eq:thetaImpact} is linear in $\dot\theta$, it is clear that the impact map for $\dot\theta^2$ is also linear in $\dot\theta^2$. Since compositions of affine functions remain affine, this in turn implies that a similar (affine) formular holds for $\dot\theta^2$ post-impact, and even after several impacts.

For the class of systems we consider, the total mechanical energy of the system is given by
\[
H(q,\dot q) = \dot q\trn M(q)\dot q+V(q).
\]
When the system is operating under virtual constraints, this reduces to
\[
\bar H(\theta, \dot\theta) := \underbrace{\Phi'(\theta)\trn M(\Phi(\theta))\Phi'(\theta)}_{\Upsilon (\theta)}\dot\theta^2 +\underbrace{V(\Phi(\theta))}_{\Xi(\theta)},
\]
which is, for any fixed $\theta$, an affine function of $\dot\theta^2$.

Thus, the total energy at any given $\theta$ throughout the multi-step trajectory can also be obtained in closed form as an affine function of the initial $\dot\theta^2$. That is,
\begin{equation}\label{eq:energyAffine}
H(\theta, \dot\theta) = \Upsilon(\theta)\Gamma(\theta,\theta_0)\dot\theta_0^2+\Upsilon(\theta)\Psi(\theta,\theta_0)+\Xi(\theta).
\end{equation}

To summarise, the critical fact is that knowing the current $\dot\theta_0$, and assuming that the virtual constraints will be perfectly regulated, the values of $\dot\theta^2$ can be computed for any future value of $\theta$. Note that for the planned motion to be completed, it is necessary that $\dot\theta>0$ for all $\theta$, we return to this in Section \ref{sec:critical}.

\subsection{Instantaneous Controllability}

Computation of the partial closed-form solution is simplified if $\alpha(\theta)\ne 0, \ \forall \ \theta \in [\theta_0, \theta_f]$.
For just this subsection, let us defining generalized momentum as
$
p :=\frac{\partial L(q,\dot q)}{\partial \dot q}  = M(q)\dot q
$
and note that the equations of motion satisfy $\dot p = f(q,p)+B(q)u$ where $f(q,p)$ is the ``natural'' flow of the system, given by $-\partial H(p,q)/\partial q$, where $H(p,q)$ is the Hamiltonian.

A motion satisfying the virtual constraints has $\dot q = \Phi'(\theta)\dot\theta$, and therefore a momentum $p = M(\Phi(\theta))\Phi'(\theta)\dot\theta$. Now the condition
$
B^\perp(\Phi(\theta)) M(\Phi(\theta))\Phi'(\theta) = 0
$
implies that the direction in which force cannot be applied is orthogonal to the momentum at that point. This can be considered a lack of local instantaneous controllability, i.e. the possibility of adjusting momentum transversally to the constraint surface. This is a stronger notion than stabilizability, and is not strictly necessary, but it simplifies calculations and control design. A closely related condition  was studied in  \cite{plestan2003stable} regarding invertability of coupling matrix in partial feedback linearization. A less conservative condition could be based on the region-of-stability analysis in \cite{Manchester11a}.

\subsection{Critical Points and Motion Completion}\label{sec:critical}

Many algorithms for planning motions of fully actuated robots decompose into two stages: planning a collision-free path $q^\star(s), s\in [0, S]$, and then planning or realizing a dynamically feasible time parameterization $s^\star:[0, T]\rightarrow [0,S]$, giving the feasible trajectory $q^\star(s^\star(t))$. For underactuation degree one systems and virtual constraints, the situation is similar but with one key difference: the virtual constraint represents a collision-free path in $\Q$, but given a particular virtual constraint and initial condition $\theta, \dot\theta$, the time evolution of the system is then {\em fixed} by \eqref{eq:abg}.

This is analogous to a bead sliding (without friction) along a curved wire of finite length: the path through three-dimensional space is fixed, but depending on the initial velocity the bead may complete the path, or it may stop at some point and slide back along its path, either returning to its start point or oscillating in a potential well. In the case of an underactuated walking robot controlled by virtual constraints, if the initial velocity is too low, the robot will not complete the footstep, but will fall backwards (see, e.g., the phase portrait Figure 6 in \cite{manchester2011stable}).

Given that the continuous-phase reduced dynamics satisfy \eqref{eq:abg}, if $\gamma(\theta_c)=0$ for some $\theta_c$, then $\theta = \theta_c,\dot\theta = \ddot\theta =0$, is a feasible equilibrium solution. For typical virtual constraints corresponding to walking motions, $\gamma(\theta)$ will have a single sign change over the interval $[\theta_0, \theta_f]$, at a point correspond to the peak potential energy of the system. We assume this to be the case and denote the critical value $\theta_c$.

If $\dot\theta^2(\theta_c)>0$, then robot has positive forward motion at the point of peak potential energy, and the constraint is dynamically feasible. If $\dot\theta^2(\theta_c)=0$ then the robot approaches this critical point along a heteroclinic orbit towards the balance equilibrium. If $\dot\theta^2(\theta_c)<0$ there is no real solution for $\theta$ and this corresponds to the robot not having enough velocity to pass the critical point, and falling backwards.

\subsection{An Ordering for Sets of Virtual Constraints}\label{sec:ordering}

The fact that the partial closed form solutions for $\dot\theta^2$ is {\em affine}  in $\dot\theta_0^2$ allows us to construct an ordering of virtual constraints, by which the search for an appropriate virtual constraint is logarithmic in the number of virtual constraint primitives in the library rather than linear.

Suppose there is a particular target velocity $a$ that should be met as closely as possible by $\dot\theta$ at the critical point $\theta_c$. Now, consider that the relation
\[
\dot\theta^2(\theta_c) = \Gamma_p(\theta_c,\theta_0)\dot\theta_0^2+\Psi_p(\theta_c,\theta_0) \ge a^2
\]
implies
\[
\dot\theta_0^2\ge \frac{a^2-\Psi_p(\theta)}{\Gamma_p(\theta)},
\]
and likewise for $\le$. 
We introduce an ordering of motion primitives $\prec_a$, defined like so: given two virtual constraints $p$ and $q$, with the same $q_i$ and $q_f$, then $p \prec_a q$ if
\[
\frac{a^2-\Psi_p(\theta)}{\Gamma_p(\theta)} \le \frac{a^2-\Psi_q(\theta)}{\Gamma_q(\theta)}.
\]
Now, consider the following problem: for a known initial velocity $\theta_0$, find the virtual constraint $p$ for which $\theta_c$ is as close as possible to $a$. From the above it is clear that if a large number $P$ of primitives are stored in the order defined by $\prec_a$, then a simple binary search can be used to find the best primitive in $O(\log P)$ time, as opposed to $O(P)$ time for checking each primitive individually.

Similar orderings could be constructed based on total energy, which is also affine in $\dot\theta_0^2$, though we defer discussion on that for a later work.


\section{Offline Construction of a Primitive Library}

In this section we describe how a library of motion primitives can be structured so as to enable real-time on-line evaluation and selection.

\subsection{Discretization of Impact Configurations}

A motion primitive $p$ is defined as a configuration path from immediately after one impact to immediately before the next. This path is parameterized by $\theta\in[\theta_0, \theta_f]$, and given by the virtual constraint $\Phi_p:[\theta_0, \theta_f]\rightarrow \Q$, with $\Phi_p(\theta_0)$ being the initial configuration and $\Phi_p(\theta_f)$ being the final configuration. Note that for different primitives, the numerical values of $\theta_0$ and $\theta_f$ may be different.

We suppose a finite library of impact configurations has been constructed, denoted by $\tilde Q$. For every primitive $p$, both $\Phi_p(\theta_0)$ and $\Phi_p(\theta_f)$ must be elements of $\tilde Q$. Since each element of $\tilde Q$ corresponds to an impact configuration, each has a particular step length $x_f$ and step height $y_f$, corresponding to the relative positions of the back (previous stance) and front (next stance) feet. For the compass-gait walker, the full configuration is completely determined by $x_f$ and $y_f$, but for robots with more degrees of freedom this will not be the case

We construct a finite set $X_f$ containing $n_x$ individual values of $x_f$, similarly $Y_f$ contains $n_y$ individual values of $y_f$, and a set $Q_o$ containing $n_q$ individual configurations of other joints, when present. Thus the cardinality of the set of impact configurations is $|\tilde Q|=n_xn_yn_q$.

\subsection{The Library of Motion Primitives}

A motion primitive connects one element in $\tilde Q$ to another. During the continuous phase, there is the opportunity to increase or decrease total energy in the system, as well as plan motions that have different trade-off between energy efficiency and collision avoidance, e.g. different bending of the swing-leg knee, if present. For each pair $\Phi_p(\theta_0),\Phi_p(\theta_f)\in \tilde Q \times \tilde Q$, we construct are $n_p$ different smooth paths $\Phi_p(\theta), \theta\in(\theta_0, \theta_f)$. Each should of course be kinematically feasible for the robot (e.g. free of self-collisions or over-extensions of joints).

We build the library of primitives in a hierarchical structure that will aid rapid search on-line. At the root are the $|\tilde Q| $ initial configurations $\Phi_p(\theta_0)\in \tilde Q$. For each such configuration there are $n_x$ step lengths $x_f$. For each pair $(\Phi_p(\theta_0), x_f)$ there are $n_y$ step heights $y_f$.

Now, for each triple $(\Phi_p(\theta_0), x_f, y_f)$ there are $n_qn_p$ virtual constraints that start with configuration $\Phi_p(\theta_0)$ and take a step of length $x_f$ and height $y_f$. We construct a balanced binary search tree of the $n_qn_p$ virtual constraints using the ordering in Section \ref{sec:ordering}. The elements of the tree are pointers to a data structure containing information about the virtual constraint primitive, detailed in the next section. We denote this tree by BST$(\Phi_p(\theta_0), x_f, y_f)$. The function TREE-SEARCH($T,a^2$) returns the primitive from the tree $T$ with $\dot\theta^2(\theta_c)$ as small as possible subject to $\dot\theta^2(\theta_c)\ge a^2$.

\subsection{Data Stored For Each Primitive}

For each motion primitive $p$, we store the affine functions for evaluation of $\dot\theta^2$ given $\dot\theta^2_0$ at the critical point ($\Gamma(\theta_c, \theta_0), \Psi(\theta_c, \theta_0)$), immediately before impact ($\Gamma(\theta_f, \theta_0), \Psi(\theta_f, \theta_0)$), and immediately post-impact ($\Gamma(\theta_p, \theta_0), \Psi(\theta_p, \theta_0)$).

For the idealized case of planar bipedal walking over uneven terrain, in which the ``world'' consists only of the robot and the ground terrain, the problem of collision checking is quite simple. For each virtual constraint $p$ there exists an envelope function $\bar q_p:[x_i, x_f]\rightarrow \R$ such that every physical point on the robot $(x,y)$ satisfies $y\ge \bar q_p(x)$. For typical walking robots and motions, $\bar q_p(x)$ will be the arc swept out by the lowest point on the swing leg.

\section{The Motion Planning Algorithms}


%
The purpose of the on-line algorithm is to find a feasible sequence of virtual constraints. The algorithms we propose can be understood as iteratively searching through a decision tree, in which nodes correspond to robot impact states (configurations and velocities), and edges correspond to virtual constraint primitives linking them. The root node is the robot's current state, and the ``depth'' in the tree corresponds to the number of footsteps ahead being planned.

If the library contains $l$ feasible primitives for each step, and there are $k$ footsteps to be planned, then an exhaustive search must evaluate $l^k$ possibilities. The objective is to greatly reduce the number of paths through the tree that are evaluated. The algorithms we suggest use a form of ``best first search'', but they differ by how ``best'' is evaluated: either trying to simply attain a particular critical velocity $\dot\theta(\theta_c)\approx a$ or using heuristics based on predicted energy requirements.
%
%
%
%
%
%
%

\subsection{Real-Time Selection and Evaluation of Primitives}

Unless the environment is purpose made (e.g. in a factory), it is unlikely that the terrain exactly matches the quantized heights chosen. In this paper, we introduce a generic function QUANTIZE$:\R \rightarrow \{Y_f, \emptyset\}$ that  maps a height value from a terrain map to the closest element of $Y_f$ or return a null value $\emptyset$ if there is no element of $Y_f$ sufficiently close. We assume that errors between the exact terrain and the quantization introduced in motion planning can be tolerated by the feedback control mechanism.

The first constraint on dynamic feasibility is that the robot has enough kinetic energy to complete the step. Since the planned motion assumes $\theta$ to be monotonically increasing, this corresponds to the constraint that
$\dot\theta>0$ for the duration of the motion. That is, the robot does not stop part way through and fall backwards. This can be enforced by ensuring $\dot\theta^2(\theta_c)>0$.

We assume a perfectly inelastic collision, i.e. no bouncing or slipping when the swing foot becomes the stance foot. Many real materials exhibit velocity-dependent coefficients of restitution \cite{stewart2000rigid}, so to ensure there is no slipping or bouncing it may be necessary to impose a bound on velocity at impact: $|\dot\theta|<b$ when $q\in \Sw$. This can obviously be converted into the bound $\dot\theta^2(\theta_f)<b^2$.

The virtual constraint $p$ is then feasible if $\bar q_p(x)\ge h(x)$ for all $x\in[x_i, x_f]$. Since these functions are one-dimensional, checking this on sufficiently fine gridding of $[x_i, x_f]$ will generally be acceptable and very fast to compute.

\subsection{Best-First Search}\label{sec:bfs}

\begin{figure}
\begin{tikzpicture}[grow'=right,level distance=0.9in,sibling distance=.05in]
\tikzset{edge from parent/.style= 
            {thick, draw, edge from parent fork right},
         every tree node/.style=
            {draw,minimum width=0.3in,text width=0.7in}}
\Tree 
    [.\node[draw,fill=white!80!green, text width=0.45in]{Current State $x$};
        [.\node[draw,fill=white!80!green]{$x.p_a$\hfill 4};
            [.{$x.p_a.p_d$} ]
            [.\node[draw,fill=white!80!green]{$x.p_a.p_e$\hfill 6 }; [.{$x.p_a.p_e.p_j$} ] [.\node[draw,fill=white!80!green]{$x.p_a.p_e.p_k$\hfill 7}; ] [.{$x.p_a.p_e.p_l$} ]]
            [.\node[draw,fill=white!60!red]{$x.p_a.p_f$\hfill 5};  ]
        ]
        [.\node[draw,fill=white!60!red]{$x.p_b$\hfill 1}; 
            [.\node[draw,fill=white!60!red]{$x.p_b.p_g$\hfill 3}; ]
            [.\node[draw,fill=white!60!red]{$x.p_b.p_h$\hfill 2}; ]
            [.{$x.p_b.p_i$} ]
        ] 
        [.{$x.p_c$} ]
    ]
\end{tikzpicture}
\caption{Illustration of a hypothetical motion primitive tree, for planning three footsteps ahead with three primitives available at each node. Green and red mark nodes that are ruled feasible and infeasible, respectively, while white nodes are never evaluated. The numbers on the right show the sequence which primitives are evaluated. See discussion in Section \ref{sec:bfs}.}
\label{fig:tree}
\end{figure}
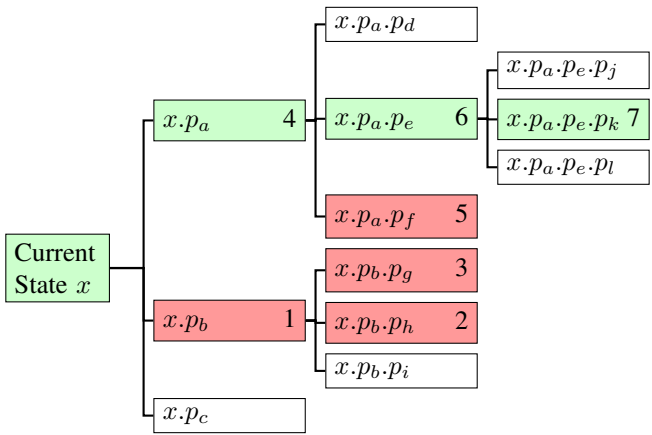

A full listing is given in Algorithm \ref{mpalg}. Here we explain the reasoning behind the algorithm by way of a hypothetical tree shown in Figure \ref{fig:tree}. We refer to algorithm line $x$ by (L$x$).

At initialisation, $q, \dot q$ and $h(x)$ are given, and $k=3$ is the number of footsteps to plan. At the first call of ADD-NODE (L\ref{alg:firstadd}), motion primitives $p_a, p_b, p_c$ are available. For a particular step length $x_f$, the TREE-SEARCH function returns $p_b$ (L\ref{alg:treesearch}), and is found to have feasible final velocity and be free of collisions (L\ref{alg:feas1}-\ref{alg:feas}). It's post-impact value of  $\dot\theta^2(\theta_c)$ is recorded as $v_c$ (L\ref{alg:vc}).

This is repeated for each step length (L\ref{alg:length4}-\ref{alg:lengthEnd4}). By construction, all primitives have $v_c\ge a^2$, so the primitive with the smallest $v_c$ is selected (L\ref{alg:argmin}), its post-impact state is computed (L\ref{alg:finalstate1}-\ref{alg:finalstate2}), and ADD-NODE is called with the footstep count decremented (L\ref{alg:newadd}).

On this second call of ADD-NODE, the TREE-SEARCH returns $p_h$, which is found to be infeasible (L\ref{alg:feas1}-\ref{alg:feas}), so $v_c$, representing $\dot\theta^2(\theta_c)$ is set to $\infty$. The algorithm next tries the successor of $p_h$, which is $p_g$ (L\ref{alg:succ}). This is also found to be infeasible, and exhausts the list of possibilities (L\ref{alg:nullsucc}), so ADD-NODE returns failure.

\begin{algorithm}
  \caption{BEST-FIRST-SEARCH($q, \dot q, h(\cdot), k$)}
  \label{mpalg}
  \begin{algorithmic}[1]
    \State [$p$, status]$\leftarrow$ ADD-NODE$(q, \dot q, h(x), k)$   \label{alg:firstadd}
    \If{status = success}
     	\State MotionControl $\leftarrow p$ \label{alg:control}
    		\State \Return success
    	\Else
    		\State \Return fail
    	\EndIf
    
  \Statex
  \Function{ADD-NODE}{$q, \dot q, h(x)$}
  \ForAll{$x_f \in X_f$}
  \State $y_f \leftarrow $ QUANTIZE$(h(x_f))$
  \State $p(x_f) \leftarrow $ TREE-SEARCH(BST$(q, x_f, y_f), a^2)$ \label{alg:treesearch}
  \EndFor
  
  \While{true}
  \ForAll{$x_f \in X_f$}\label{alg:length4}
  \State $v_0 \leftarrow (\Phi_{p(x_f)}'(\theta_0)\dot q)^2$\label{alg:feas1}
  \State $v_f \leftarrow \Gamma_{p(x_f)}(\theta_f)v_0+\Psi_{p(x_f)}(\theta_f)$
  \If{$v_f\le b^2$ AND $q_{p(x_f)}>h(x)$} \label{alg:feas}
  	\State $v_c(x_f) \leftarrow \Gamma_{p(x_f)}(\theta_c)v_0+\Psi_{p(x_f)}(\theta_c)$ \label{alg:vc}
  \Else
    \State $v_c(x_f) \leftarrow \infty$
    \If {$p(x_f).$successor $\ne \varnothing$}
  	  \State $p(x_f)\leftarrow p(x_f).$successor \label{alg:succ}
  	\EndIf
  \EndIf
  \EndFor\label{alg:lengthEnd4}
  
  \State $v^\star=\min_{x_f\in X_f}(v_c(x_f))$
  \If{$v^\star \ne \infty$}
  	\State $x^\star=\arg\min_{x_f\in X_f}(v_c(x_f))$ \label{alg:argmin}
  	\If {$k=1$}
  		\State \Return [$p(x^\star)$, success] \label{alg:success}
  	\EndIf
  	\State $q^+ \leftarrow R\Phi_{p(x^\star)}(\theta_f)$\label{alg:finalstate1}
  	\State $\dot q^+ \leftarrow \Phi_{p(x^\star)}'\sqrt{\Gamma_{p(x^\star)}(\theta_p)v_0+\Psi_{p(x^\star)}(\theta_p)}$ \label{alg:finalstate2}
  	\State $k\leftarrow k-1$
  	\State [p,status] $\leftarrow$ ADD-NODE$(q^+, \dot q^+, h(x),k)$	\label{alg:newadd}
  	\If {status $=$ success} \label{alg:statuscheck}
  		\State \Return [$p(x^\star)$, success] \label{alg:backprop}
    \Else
    		\State $p(x^\star) = p(x^\star).$successor 
  	\EndIf  	
  	
  \EndIf
  
  \If{$\forall x_f\in X_f, p(x_f)$.successor $= \varnothing$}\label{alg:nullsucc}
  	\State \Return fail
  \EndIf	  
  \EndWhile  
  
  \EndFunction
      
\end{algorithmic}
\end{algorithm}

At this point, program flow returns to (L\ref{alg:statuscheck}) for the first call of ADD-NODE with a fail status. Hence the successor to $p_b$ is chosen, which is $p_a$, which is found to be feasible. Supposing TREE-SEARCH returns $p_e$ followed by $p_k$, each of which is feasible, and each of which corresponds to a call of ADD-NODE with $k$ decremented, then $k$ has reduced to 1 and the algorithm returns success (L\ref{alg:success}).

The algorithm has now found a feasible three-step sequence of primitives, so this success status propagates back up the tree (L\ref{alg:backprop}) so the primitive $p_a$ is sent to the motion control system (L\ref{alg:control}).

%
%
%
%
%

\subsection{Improvement via Energy Heuristics}

The obvious disadvantage of the myopic {\em best-first} algorithm is the inability to select the motion primitives best suited to changes in forthcoming terrain, particularly deviations in altitude. An improved strategy involves {\em looking ahead} a finite distance to estimate the net change in altitude over this range, providing an approximation of the corresponding change in gravitational potential energy, which is equivalent to the additional energy that we must accumulate (or dissipate) if we are to maintain the current kinetic energy. Dividing this net change in energy by the approximate number of footsteps gives the required incremental energy change per footstep, which may then be compared with the {\em post-impact} change in energy resulting from a possible footstep to assess its suitability. We constructed a similar algorithm -- details ommitted due to space restrictions -- making use of orderings of primitives in terms of the affine solution of total energy \eqref{eq:energyAffine}.

%

\section{Simulation Results}

This section contains results of simulations using the proposed algorithms to plan motions over uneven terrain for two frequently-studied models of walking robots: the compass gait walker, and a five-link walker. The dynamical models for both were taken from \cite{westervelt2007feedback}. The matlab code for our algorithms and simulations on these models can be found at the author's website \cite{software}.

%

%
%

The compass-gait walker is a simple two-degree-of-freedom planar biped consisting of two straight, rigid legs which meet at a revolute joint, called the {\em hip}, where a single actuator can apply a torque. In practice, such a robot is fitted with retractable pointed feet to overcome the problem of ``toe-scuffing'' due to the absence of knee joints; in this study, foot retractions are not modeled as their mass is considered negligible. 

%

\begin{figure}
\begin{center}
\includegraphics[width=1\columnwidth]{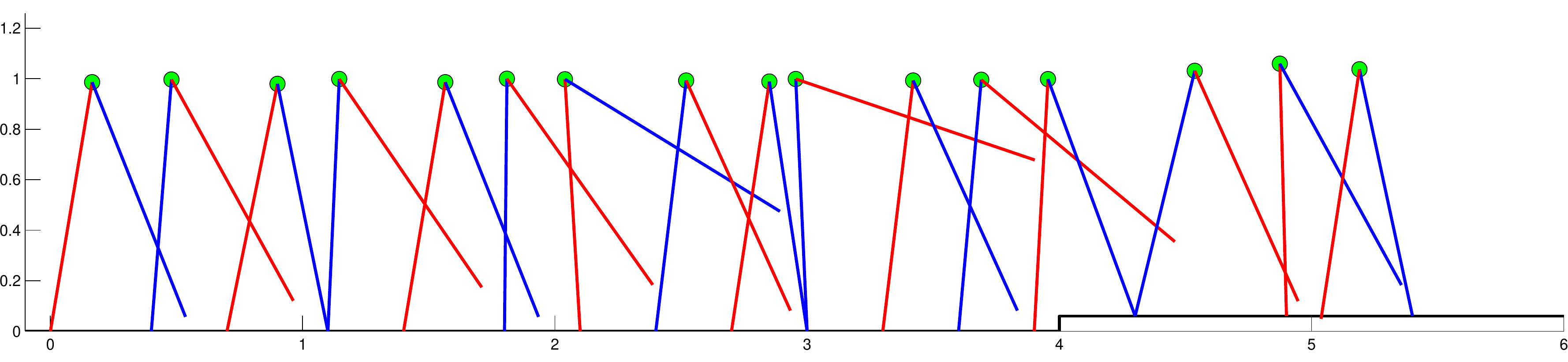}
\caption{A step-up trajectory of the Compass Gait walker.}
\label{fig:cgPath1}
\end{center}
\end{figure}

\begin{figure}
\begin{center}
\includegraphics[width=0.8\columnwidth, trim = 105pt 270pt 95pt 280pt, clip=true]{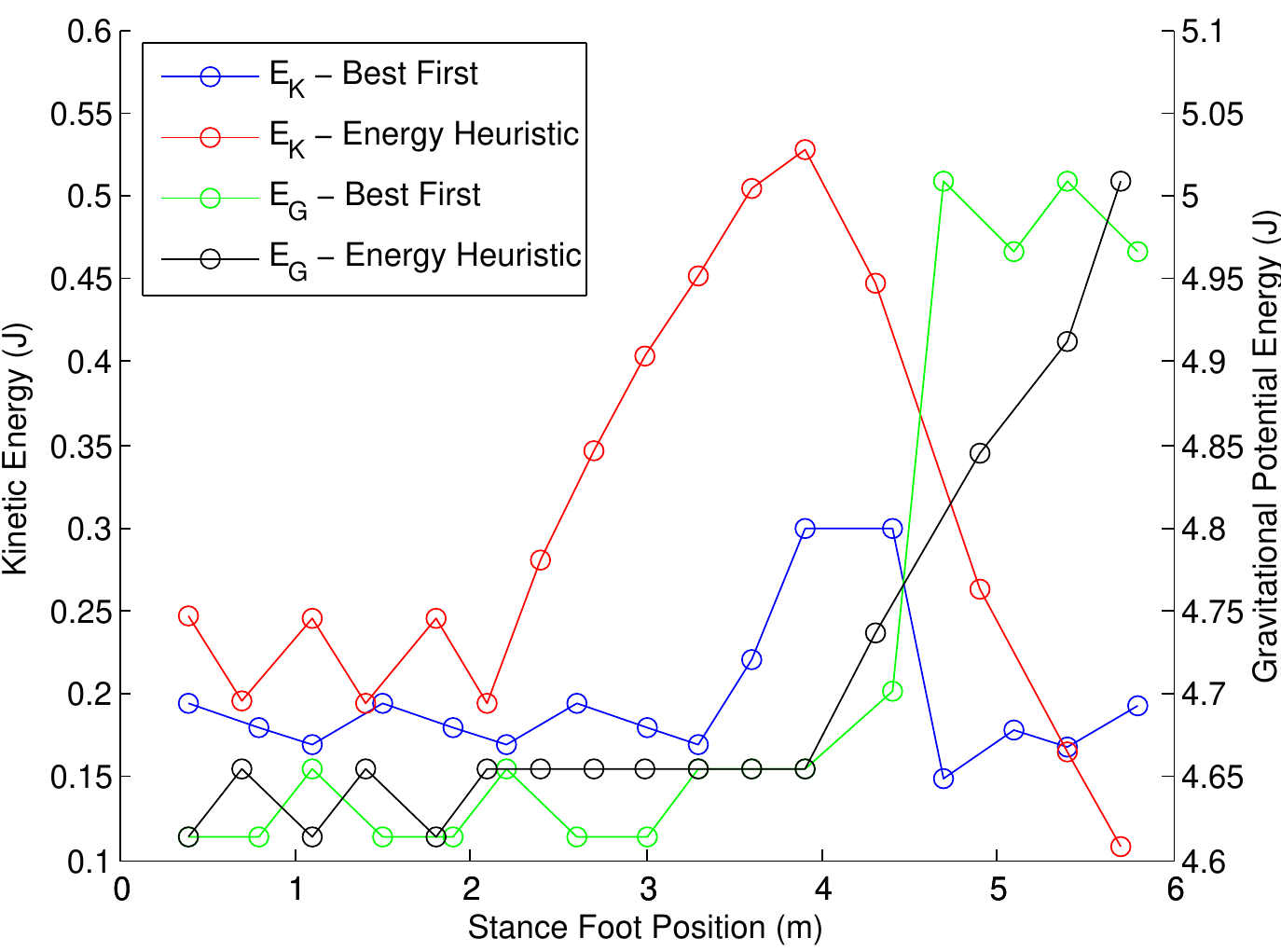}
\caption{Kinetic and potential energy changes with best-first search and
  the energy heuristic for the step-up trajectory in Figure \ref{fig:cgPath1}.}
\label{fig:cgEnergy}
\end{center}
\end{figure}

In the Figure \ref{fig:cgPath1} we show a simple step-up trajectory performed with a five-step lookahead. We can see that the walker swings its leg high in the lead up to the step, which puts the centre of mass further in front of the pivot point and adds energy. Figure \ref{fig:cgEnergy} shows the kinetic and potential energy of the walker for both algorithms for this terrain. We can see that with both algorithms there is a build-up of kinetic energy before the step-up, which then transfers to potential energy. However, with the energy heuristic the build-up begins earlier and more accurately approaches the required energy. The build-up from the greedy best-first search occurs simply because otherwise the trajectories would be infeasible.

\begin{figure}
\begin{center}
\includegraphics[width=1\columnwidth]{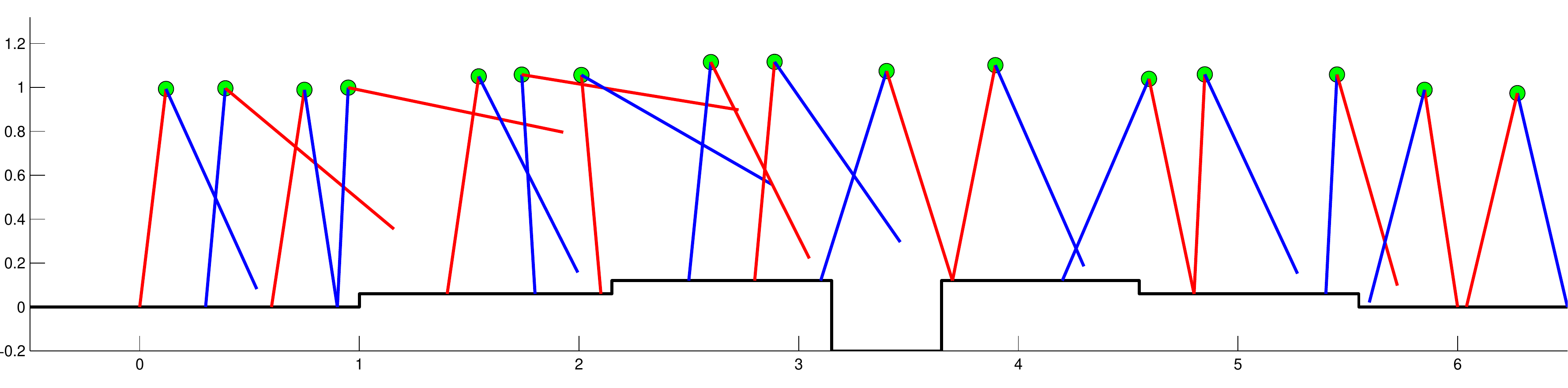}
\caption{A trajectory of the Compass Gait walker over more varied terrain.}
\label{fig:cgPath2}
\end{center}
\end{figure}

In Figure \ref{fig:cgPath2} we show another slightly more difficult terrain, with several steps up and down, and a gap. We can again see the walker kicks its leg forward to build up energy for the steps up, and keeps it low for the steps down. 

\begin{figure}
\begin{center}
\includegraphics[width=0.8\columnwidth]{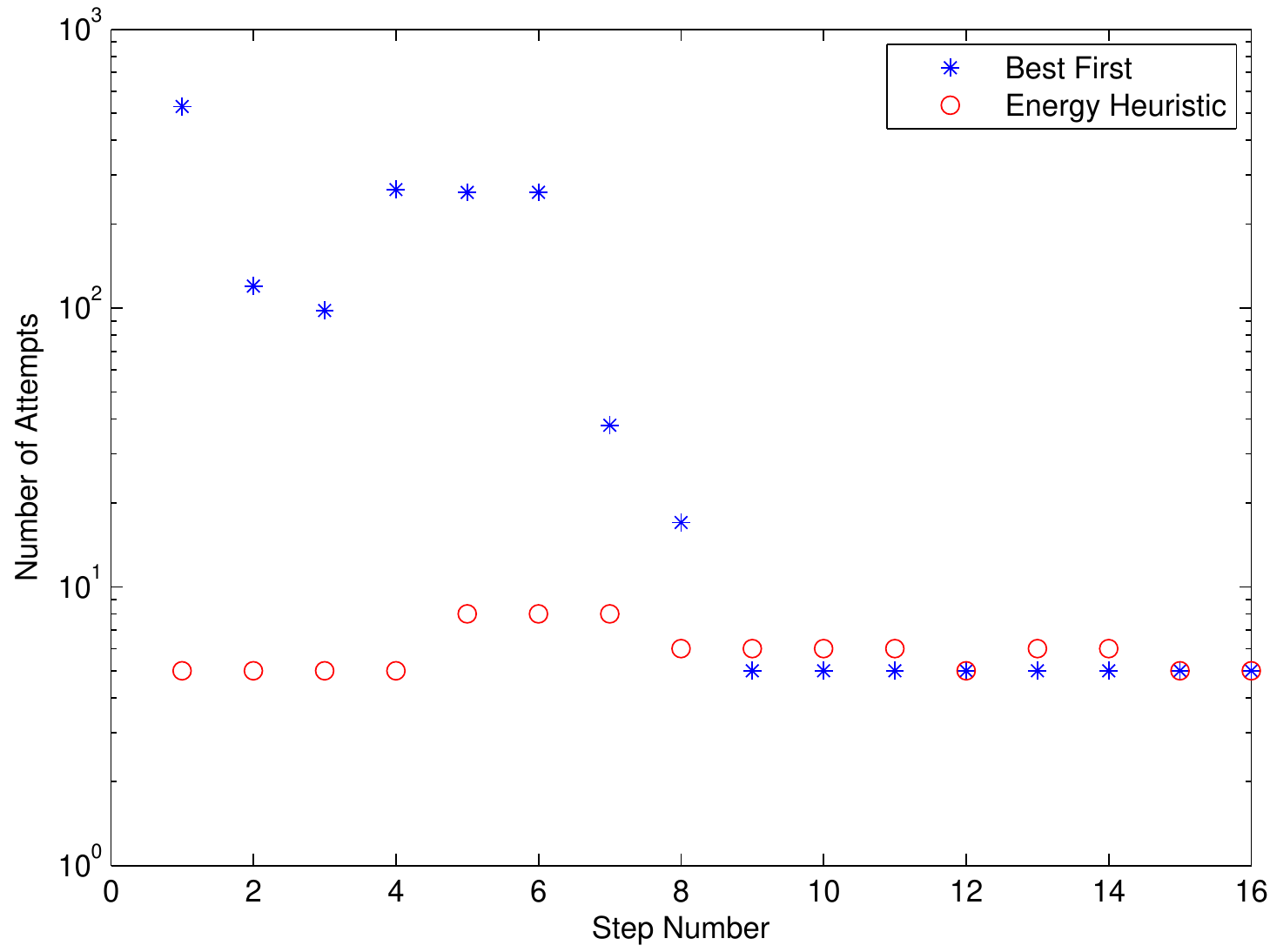}
\caption{Number of nodes evaluated to plan the trajectory in Figure
  \ref{fig:cgPath2}, for best-first search and the energy heuristic.}
\label{fig:cgAttempts}
\end{center}
\end{figure}

For this simulation, there were 24 possible virtual constraints for each step, and a five-step lookahead. This implies the total search tree has $24^5$ -- about 8 million -- possible trajectories for each plan. Figure \ref{fig:cgAttempts} shows the benefit of the energy heuristic in terms of computation time. It depicts the number of nodes evaluated, i.e. the number of times the ADD-NODE function in Algorithm \ref{mpalg} was run. The worst case scenario is that it is run 8 million times, the best case scenario is that it chooses nodes perfectly, so it is run five times (one for each footstep). As we can see, for the more difficult terrain the best-first search evaluated several hundred possible nodes, whereas the energy heuristic algorithm always evaluated less than ten. Since each evaluation requires only a small number of arithmetic operations, both algorithms could easily be run in real time on a low-power microcontroller.

We also applied the algorithm to the more complex 7 degree-of-freedom/5-link walker from \cite{plestan2003stable}, modeled on the Rabbit robot \cite{chevallereau2003rabbit}.  Each leg consists of two rigid links connected by a knee joint, while an additional rigid link - the torso - also extends from the hip joint. The joint between each femur and the torso is actuated, as are both knees, so an independent control torque can be applied to each of these four joints.
%

\begin{figure}
\begin{center}
\includegraphics[width=1\columnwidth]{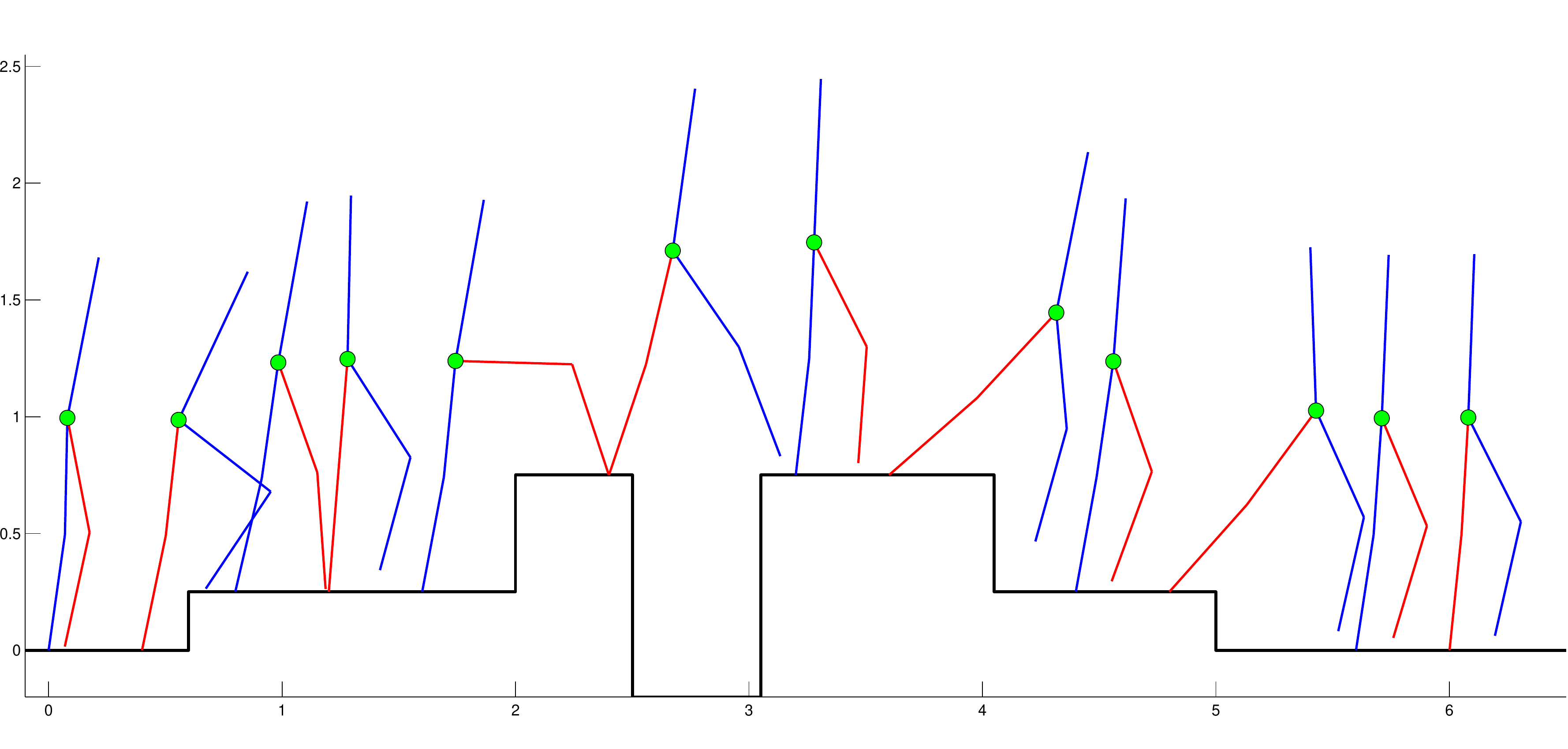}
\caption{A trajectory of the five-link walker over uneven terrain.}
\label{fig:fivelinkPath}
\end{center}
\end{figure}

In Figure \ref{fig:fivelinkPath} we show the walker negotiating quite difficult terrain, with large steps up and down and a gap to step across. Notice that the walker leans its torso forwards for the early steps to increase energy for the steps up, and leans it back to slow down for the steps down.

%

%
%
%
%
%
%
\bibliographystyle{IEEEtran}

\bibliography{motionplanning}

\end{document}